\documentclass[12pt]{article}
\usepackage{epsfig}

\def \pq  {\mathop{\rm pq}\nolimits}

\def \ds  {\mathop{\rm ds}\nolimits}
\def \ns  {\mathop{\rm ns}\nolimits}
\def \cs  {\mathop{\rm cs}\nolimits}
\def \sn  {\mathop{\rm sn}\nolimits}
\def \cn  {\mathop{\rm cn}\nolimits}
\def \dn  {\mathop{\rm dn}\nolimits}
\def \baru {\overline{u}}
\def \barU {\overline{U}}

\def \ha  {\mathop{\rm h}\nolimits}

\def \mod#1{\vert #1 \vert}
\def \sech{\mathop{\rm sech}\nolimits}
\def \coth{\mathop{\rm coth}\nolimits}

\def \ax {a_x} \def \at {a_t} \def \kx {k_x} \def \kt {k_t}
\def \ax {r}   \def \at {s}   \def \kx {k}   \def \kt {k_1}
   
\begin{document}

\title{Periodic waves of a discrete higher order
nonlinear Schr\"odinger equation\footnote
{
Communications in theoretical physics, to appear.
Corresponding author R.~Conte, phone +33--169087349, fax +33--169088786.
nlin.PS/0604010. Preprint S2005/105.} 
}

\author{Robert Conte\dag,\ K.W.~Chow\ddag\
{}\\
\\ \dag Service de physique de l'\'etat condens\'e (URA 2464), CEA--Saclay
\\ F--91191 Gif-sur-Yvette Cedex, France
\\ E-mail:  Robert.Conte@cea.fr
{}\\
\\ \ddag Department of Mechanical Engineering, University of Hong Kong
\\ Pokfulam, Hong Kong
\\ E-mail: KwChow@hkusua.hku.hk
{}\\
}

\maketitle

{\vglue -10.0 truemm}
{\vskip -10.0 truemm}

\begin{abstract}
The Hirota equation is a higher order extension of the nonlinear
Schr\"odinger equation by incorporating third order dispersion and
one form of self steepening effect.
New periodic waves for the discrete
Hirota equation are given in terms of elliptic functions.
The continuum limit converges to the analogous result for the continuous
Hirota equation,
while the long wave limit of these discrete periodic
patterns reproduces the known result of the integrable Ablowitz-Ladik
system.
\end{abstract}



\noindent \textit{PACS 2001} 
  02.30.Jr,
  02.30.Ik,
  42.65.Wi.


\baselineskip=12truept


\tableofcontents

\section{Introduction}

\def \ax {a_x} \def \at {a_t} \def \kx {k_x} \def \kt {k_t}
\def \ax {r}   \def \at {s}   \def \kx {k}   \def \kt {k_1}
\indent
\phantom{xx}
\indent

The Hirota equation \cite{H1973a}
\begin{eqnarray}
& &
 i u_t +   c_1 \left(p u_{xx} +   q {\mod u}^2 u\right)
       + i c_2 \left(p u_{xxx}+ 3 q {\mod u}^2 u_x\right)=0,\
i^2=-1,\ c_2 p q \not=0,
\label{eqhNLS_continuous}
\end{eqnarray}
in which $p,q,c_1,c_2$ are real constants
and $u$ is a complex function of $x$ and $t$,
is quite important in the propagation of short pulses in optical fibers
and lattice dynamics \cite{KT1996,XuLiLiZhou,LiLiLiZhou}.
Indeed,
this is one of the only two integrable cases
\cite{SasaSatsuma,Sakovich1997}
of a class of equations \cite{K1985,KH}
\begin{eqnarray}
& &
 i u_t + \alpha_1 u_{xx} + \alpha_2 {\mod u}^2 u
 + i \left(
\beta_1 u_{xxx} + \beta_2 {\mod u}^2 u_x +\beta_3 \left({\mod u}^2\right)_x u
\right)=0,\
(\alpha_j,\beta_j) \hbox{ real},
\label{eqhNLSgeneral}
\end{eqnarray}
which incorporates the higher-order correction terms
needed when the nonlinear Schr\"odinger equation (NLS)
becomes insufficient
or inadequate to describe the relevant physical phenomena.

Just like the NLS,
equation (\ref{eqhNLS_continuous}) is invariant under a Galilean
transformation \cite{Sakovich1997,GHNO} which could allow one to set
$c_1=0,c_2=1$
without loss of generality.
It is however preferable to retain both parameters $c_1,c_2$
in order to better display the two limits,
\begin{description}
\item (a)
the nonlinear Schr\"odinger equation $(c_1=1,c_2=0)$,
\begin{eqnarray}
& &
 i u_t +   p u_{xx} +   q {\mod u}^2 u=0,
\label{eqNLS_continuous}
\end{eqnarray}

\item (b)
the modified Korteweg-de Vries equation (mKdV) $(c_1=0,c_2=1,\baru=u)$,
\begin{eqnarray}
& &
 u_t + p u_{xxx}+ 3 q u^2 u_x=0.
\label{eqmKdV_continuous}
\end{eqnarray}

\end{description}

If the time $t$ is kept continuous and only the space variable $x$ is made
discrete,
one can build a discretization of (\ref{eqhNLS_continuous})
which admits a discrete Lax pair \cite{PL1989}
and therefore remains integrable.
With the notation $U_n = U(X, t),\ X= n h$,
this semi-discrete equation is
\begin{eqnarray}
& &
i U_{n,t}
+ \frac{1}{2}\left\lbrack
   \nu (U_{n+1}+U_{n-1}) + i \mu(U_{n+1}-U_{n-1})\right\rbrack
   (1+\delta U_n \barU_n) - \nu U_n = 0,\
\label{eqhNLS_discrete}
\end{eqnarray}
and its continuum limit to (\ref{eqhNLS_continuous})
involves a convenient linear change of coordinates
aimed at avoiding a $U_{n,x}$ term in
the definition of (\ref{eqhNLS_discrete}),
\begin{eqnarray}
& & \left\lbrace
\begin{array}{ll}
\displaystyle{
U_n=U(X,t)=u(x,t),\ X=n h = x + c_0 t,
}
\\
\displaystyle{
\delta=\frac{q}{2 p} h^2,\
\nu=2 c_1 p h^{-2},\
\mu=6 c_2 p h^{-3},\
c_0=6 c_2 p h^{-2}.
}
\end{array}
\right.
\label{eqhNLS_discrete_continuum_limit}
\end{eqnarray}

 Bright and dark solitary (or localized) waves will occur when the
 signs of the dispersive and nonlinear terms in the NLS portion of
 (\ref{eqhNLS_discrete})
 are of the same and opposite signs respectively.
 The explicit expressions have been given earlier \cite{N1990,N1991dHNLS}.
 In particular,
 bright and rational solitons on a continuous wave background are also obtained
 for the case of dispersive and nonlinear terms of the same sign.

 The main goal of the present paper
 is to deduce new, special \textit{periodic}
 solutions of (\ref{eqhNLS_discrete}).
 First we shall start with solutions of the continuous Hirota equation
 (\ref{eqhNLS_continuous}),
 some of which, surprisingly, have not yet been documented.
 The discrete analogue of these solutions will yield
 exact results for (\ref{eqhNLS_discrete}).

\section{Continuum and discrete one-soliton solutions}
\label{sectionOne-soliton_solutions}

We begin by making a rather puzzling remark on localized
pulses of these higher order envelope equations.
In the defocusing regime ($p q <0$),
where conventional wisdom will dictate the presence of dark solitons,
a localized pulse which rises above the
mean level of a continuous background can exist.
This solution,
\begin{eqnarray}
& & \left\lbrace
\begin{array}{ll}
\displaystyle{
u=e^{i(K x - \omega t)}
 a_0 \left\lbrack
 \frac{k \sinh k a}{\cosh k (x-ct) + \cosh k a} + i Y \right\rbrack,\
 a_0^2=- \frac{2 p}{q},\
}
\\
\displaystyle{
 K=\frac{c_1}{3 c_2},\
}
\\
\displaystyle{
\omega=c_1 p \left(2 Y^2 -\frac{k^2}{3}+\frac{k^2 \coth^2 ka}{2}\right)
      -c_2 p K^3,\
}
\\
\displaystyle{
c=c_2 p \left(k^2 -\frac{3}{2} (k \coth ka)^2 + 3 K^2 - 6 Y^2\right),\
\hbox{arbitrary}=(a,k,Y).
}
\end{array}
\right.
\label{eqhNLS_continuous_sechm}
\end{eqnarray}
can be obtained with the method in \cite[App.~A]{CM1993}.
A discrete equivalent of (\ref{eqhNLS_continuous_sechm})
is not yet known.

\section{Elliptic travelling wave solution in the continuum case}
\label{sectionContinuum_travelling_waves}

Doubly periodic solutions of the continuous Hirota equation
(\ref{eqhNLS_continuous}) have been documented earlier
\cite{Fan2002,Feng2003},
they are analytically identical to those \cite{Chow2000}
of the Sasa-Satsuma case \cite{SasaSatsuma} of the integrable higher NLS
(i.e.~$3 \beta_1 \alpha_2=\beta_2 \alpha_1, \beta_2=2 \beta_3$ in
(\ref{eqhNLSgeneral})).
Most of these
solutions can be expressed as either a purely real elliptic function,
mulitplied by the sinusoidal (or exponential) phase of wave oscillations.
We propose a new
form of solution expressed with nonzero real and imaginary parts.
We shall present this solution first in the notations of Halphen
(Appendix A).
Although this notation appears to be more abstract and less common,
it enjoys the advantage of displaying clearly
the ternary symmetry that exists in the complex plane
between e.g.~the three Jacobi functions $(\cn,\dn,\sn)$.
We then employ the more conventional Jacobi
elliptic functions.

Assuming a solution of (\ref{eqhNLS_continuous})
in the form of a traveling wave (see Appendix A),
\begin{eqnarray}
& &
u=a_1 \left\lbrack
 \ha_\alpha(x-ct) + i b_1 \frac{\ha_\beta(x-ct)}{\ha_\gamma(x-ct)}
      \right\rbrack
e^{i(K (x-ct) - \omega t)},\
\label{eqhNLS_continuous_elliptic_Halphen}
\end{eqnarray}
one finds the unique solution
\begin{eqnarray}
& & \left\lbrace
\begin{array}{ll}
\displaystyle{
a_1^2=-\frac{2 p}{q},\
} \\ \displaystyle{
\omega/p= c_1 (2 b_1^2 - 2 b_1 K - K^2 - 3 e_\alpha)
       +  c_2 (2 K(3 e_\alpha+K^2) + 6 b_1 (e_\beta-e_\gamma) + 6 K^2),\
} \\ \displaystyle{
c/p=2 c_1 (b_1+K) + 3 c_2 (e_\alpha - K^2 -2 b_1 K - 2 b_1^2),\
} \\ \displaystyle{
b_1 (e_\gamma-e_\alpha-b_1^2) (e_\gamma-e_\beta)=0,
}
\end{array}
\right.
\label{eqhNLS_continuous_elliptic_Halphen_sole}
\end{eqnarray}
which, as displayed in the last line, splits into three subcases
\begin{description}
\item
$(b_1=0)$:
the same elliptic solution \cite{Fan2002} as for the pure NLS case,

\item
$(e_\gamma-e_\alpha-b_1^2=0)$:
another elliptic solution,
different from the previous one,

\item
$(e_\gamma-e_\beta)=0$:
a trigonometric solution,
identical to the well known dark one-soliton
\begin{eqnarray}
u=a_0 \left\lbrack \frac{k}{2} \tanh \frac{k}{2}(x-ct)+ i b_1 \right\rbrack
 e^{i(K x - \omega t)}.
\label{eqhNLS_continuous_dark}
\end{eqnarray}

\end{description}
In each case, the solution depends on three arbitrary constants,
e.g. respectively,
$(e_\gamma,e_\alpha,K)$,
$(e_\alpha,b_1,K)$,
$(e_\alpha,b_1,K)$,
i.e. three less arbitrary constants
than the general travelling wave solution
whose expression is still unknown.
This lack of arbitrary constants
is responsible for the factorized form
of the last line in
(\ref{eqhNLS_continuous_elliptic_Halphen_sole}).

To the best of our knowledge,
the elliptic solution with $b_1\not=0$ (second case)
is new.

In the more familiar Jacobi notation \cite{Lawden},
this elliptic solution for $b_1 \not=0$
is a periodic oscillation of amplitude $A_1$,
\begin{eqnarray}
& &
u = A_1 \left\lbrack
 \cn (\ax (x-ct),\kx) +  i B_1 \frac{\sn (\ax (x-ct),\kx)}{\dn (\ax (x-ct),\kx)}
 \right\rbrack e^{i (K (x-ct) - \omega t)},
\end{eqnarray}
where
the wave vector $\ax$,
the Jacobi modulus $\kx$,
the velocity $c$
and
the angular frequency $\omega$
are related by
\begin{eqnarray}
& & \left\lbrace
\begin{array}{ll}
\displaystyle{
\kx^2=1-B_1^2,\
} \\ \displaystyle{
A_1^2=\frac{2 p}{q} \ax^2 (1-B_1^2),\
} \\ \displaystyle{
c/p=2 c_1 (K - \ax B_1)
     +c_2 (\ax^2 - 3 K^2 + 6 K \ax B_1 -8 \ax^2 B_1^2),\
} \\ \displaystyle{
\omega/p=c_1 (-K^2 + 2 K \ax B_1 + 4 \ax^2 B_1^2 - \ax^2)
} \\ \displaystyle{
\phantom{12345}
  +c_2 (2 K \ax^2 + 6 \ax^3 B_1 + 2 K^3 - 6 K^2 \ax B_1 - 4 K \ax^2 B_1^2),\
}
\end{array}
\right.
\label{eqhNLS_continuous_elliptic_Jacobi_sole}
\end{eqnarray}
and the three parameters $(a_x,B_1,K)$ are arbitrary.

\textit{Remark}.
The nonintegrable higher NLS equation (\ref{eqhNLSgeneral})
also admits the solution (\ref{eqhNLS_continuous_elliptic_Halphen}),
however with one less arbitray constant since $K$ is then fixed.

\section{Particular travelling waves in the discrete case}
\label{sectionDiscrete_travelling_waves}

If one assumes for (\ref{eqhNLS_discrete})
the same kind of solution as for (\ref{eqhNLS_continuous}), namely
\begin{eqnarray}
& &
U=a_1 \left\lbrack
 \ha_\alpha(X-ct) + i b_1 \frac{\ha_\beta(X-ct)}{\ha_\gamma(X-ct)}
      \right\rbrack
e^{i(K (X-ct) - \omega t)},\
\label{eqhNLS_discrete_elliptic_Halphen}
\end{eqnarray}
one finds solutions similar to those of the continuous case.
In particular, the elliptic solution with $b_1 \not=0$ is
\begin{eqnarray}
& & \left\lbrace
\begin{array}{ll}
\displaystyle{
a_1^2=-\frac{\ha_\gamma^2(h)}{\delta N_\alpha},\
e_\gamma=e_\alpha + b_1^2,\
} \\ \displaystyle{
\omega=\nu \left(1-\frac{\ha_\beta(h) \ha_\gamma^3(h)}{N_\alpha} \right)
 + \mu b_1 (e_\beta-e_\gamma) \frac{\ha_\alpha(h)}{N_\alpha},
} \\ \displaystyle{
c=\ha_\gamma(h) \frac{\mu \ha_\alpha(h) \ha_\gamma(h) + \nu b_1 \ha_\beta(h)}
  {N_\alpha},\
\hbox{arbitrary}=(e_\alpha,b_1,K),
}
\end{array}
\right.
\label{eqhNLS_discrete_elliptic_Halphen_sole}
\end{eqnarray}
in which $N_\alpha$ denotes the nonzero constant
\begin{eqnarray}
& &
N_\alpha=\ha_\alpha^4(h) - (e_\alpha-e_\beta)(e_\alpha-e_\gamma)
= 2 \ha_\alpha(h) \ha_\beta(h) \ha_\gamma(h) \ha_\alpha(2 h).
\end{eqnarray}
The corresponding expression in the Jacobi notation is
\begin{eqnarray}
& &
U = A_1 \left\lbrack
   \cn (\ax (X-ct),\kx) + i B_1 \frac{\sn (\ax (X-ct),\kx)}
                                {\dn (\ax (X-ct),\kx)}\right\rbrack
                                e^{i (K(x-ct)-\omega t)},
\label{eqhNLS_discrete_elliptic_Jacobi_sole}
\end{eqnarray}
in which the relations linking $(\ax,\kx,c,\omega,K,B_1,A_1)$
are easily deduced from
(\ref{eqhNLS_discrete_elliptic_Halphen_sole})
using the correspondence (\ref{eq_ps_ration_sigma_alpha_by_sigma}).

\section{Special limits}
\subsection{The continuum limit}

The continuum limit $h \to 0$ is easier to perform on
(\ref{eqhNLS_discrete_elliptic_Halphen_sole}).
The fixed constants $(\delta,\mu,\nu)$ behave as indicated in
(\ref{eqhNLS_discrete_continuum_limit}),
the elliptic functions of $h$ expand as
\begin{eqnarray}
& &
\ha_\alpha(h) = h^{-1} - \frac{e_\alpha}{2} h^2 + O(h^4),\
N_\alpha = h^{-4} - 2 e_\alpha h^{-2} + O(h^0),\
\nonumber
\\
& &
1-\frac{\ha_\beta(h) \ha_\gamma^3(h)}{N_\alpha}
\sim 1- \left(1-\frac{e_\beta h^2}{2}-\frac{3 e_\gamma h^2}{2}
 + \frac{e_\alpha (2 h)^2}{2}\right)
\sim \frac{5 e_\beta + 7 e_\gamma}{2} h^2,
\\
& &
\frac{\ha_\alpha(h) \ha_\gamma^2(h)}{N_\alpha}
=
h \left(1-\frac{3 e_\beta + 5 e_\gamma}{2} h^2 + O(h^4)\right),
\nonumber
\end{eqnarray}
therefore
\begin{eqnarray}
& &
c = 6 c_2 p h^{-3} h
 \left(1-\frac{3 e_\beta + 5 e_\gamma}{2} h^2 + O(h^4)\right)
+ 2 c_1 p h^{-2} b_1 h^2,
\\
& &
X-c t = x - (c- c_0) t
= x - p (2 c_1 b_1 - 3 c_2 (3 e_\beta + 5 e_\gamma)) t + O(h^2),
\nonumber
\end{eqnarray}
and the solution
(\ref{eqhNLS_discrete_elliptic_Halphen_sole})
indeed goes to the solution
(\ref{eqhNLS_continuous_elliptic_Halphen_sole}).

\subsection{The long wave limit}

In the long wave limit $\kx \to 1$,
the solution (\ref{eqhNLS_discrete_elliptic_Jacobi_sole})
goes to \cite{N1990}
\begin{eqnarray}
& & \left\lbrace
\begin{array}{ll}
\displaystyle{
U= a_2 \frac{\sin k h}{h} \sech k (X - c t) e^{i(K X - \omega t)},\
 a_2^2= \frac{2 p}{q},\
}
\\
\displaystyle{
\omega=\nu + \left(\mu \sin K h - \nu \cos K h\right) \cosh k h,\
}
\\
\displaystyle{
c=\left(\mu \cos K h + \nu \sin K h\right) \frac{\sinh k h}{k},\
\hbox{arbitrary}=(k,K).
}
\end{array}
\right.
\label{eqhNLS_discrete_long_wave_limit_Jacobi}
\label{eqhNLS_discrete_bright}
\end{eqnarray}

\section{Conclusion}

Periodic waves of the discrete Hirota equation have been generated from
their counterparts for the continuous Hirota equation.

Equation (\ref{eqhNLS_discrete})
combines the main features of the integrable discretization of NLS
\cite{AblowitzLadik,AOT,MY},
\begin{eqnarray}
& &
i \frac{\partial \psi_n}{\partial t}
   + \frac{\psi_{n+1} + \psi_{n-1} - 2 \psi_n}{h^2}
   + (\psi_{n+1} + \psi_{n-1})\psi_n \psi_n^{\star} = 0,
\label{eqNLSDiscreteAL}
\end{eqnarray}
and the discrete modified Korteweg-de Vries equation \cite{HirotamKdV}
\begin{eqnarray}
& &
\frac{\partial \psi_n}{\partial t}
   = \frac{(\psi_{n+1} - \psi_{n-1})(1+ \psi_n^2)}{h^2}.
\label{eqmKdVDiscrete}
\end{eqnarray}

The most general travelling wave solution of the continuous Hirota
equation is an hyperelliptic function of genus two,
as detailed in Appendix B.
The elliptic solution (\ref{eqhNLS_continuous_elliptic_Halphen_sole})
is a very particular case of this general traveling wave.
Future work should address the question of incorporating
the freedom of the three missing arbitrary constants
in the present elliptic solution.

\section*{Acknowledgments}

Partial financial support has been provided by the
Research Grants Council contracts HKU 7123/05E and
HKU 7184/04E.

\vfill \eject
\section*{Appendix A. The basic elliptic functions}
\label{sectionBasicEllipticFunctions}

In order to fully display the ternary symmetry of the Jacobi notation,
and consequently make the practical computations much easier,
following Halphen \cite{HalphenTraite},
we will choose the basis functions as
\begin{eqnarray}
& &
\ha_\alpha(u)=\sqrt{\wp(u)-e_\alpha},\
\alpha=1,2,3,
\label{eqHalphen_ha_sqrtwp}
\end{eqnarray}
in which $\wp$ is the Weierstrass elliptic function defined by
\begin{eqnarray}
& &
{\wp'}^2=4 (\wp-e_1) (\wp-e_2) (\wp-e_3) =4 \wp^3-g_2 \wp-g_3.
\end{eqnarray}
The link with more usual (but less symmetric) notation
such as $(\cs,\ds,\ns)$ or $(\sn,\cn,\dn)$ is
classical \cite[\S 16.10, 16.20, 18.9.11]{AbramowitzStegun},
e.g.~\cite[Chap II, (16) p.~46]{HalphenTraite},
\begin{eqnarray}
& &
 \frac{\cs(z)}{\ha_1(u)}
=\frac{\ds(z)}{\ha_2(u)}
=\frac{\ns(z)}{\ha_3(u)}
=\frac{u}{z}
= \frac{1}{\sqrt{e_1-e_3}},
\label{eq_ps_ration_sigma_alpha_by_sigma}
\end{eqnarray}
and our main results will also be displayed in this more usual notation,
familiar to the physicists.

Since the three factors $\wp-e_j,j=1,2,3$ of ${\wp'}^2$ are indistinguishable,
any solution depending on $(\ha_1,\ha_2,\ha_3)$
remains a solution after any permutation $(\alpha,\beta,\gamma)$ of $(1,2,3)$.
Therefore, \textit{in the complex plane},
a $\ha_1$ solution represents any of the twelve familiar functions $\pq$
of Jacobi (p,q among s,c,d,n).
However, \textit{on the real axis} of $x$ or of $x-ct$,
depending on the values of $(e_1,e_2,e_3)$,
less than twelve functions will be admissible.

For deriving the continuum limits,
it is useful to consider the Laurent expansion at the origin
\cite[Chap VII p.~237]{HalphenTraite}
\begin{eqnarray}
& &
\ha_\alpha(u)=
\frac{1}{u}
-e_\alpha \frac{u}{2}
+\left(g_2-5 e_\alpha^2\right) \frac{u^3}{40}
+\frac{5 g_3 -7 e_\alpha g_2}{2^6.5.7} u^5
\nonumber \\ & & \phantom{xxxxx}
+\frac{28 g_2^2 - 225 e_\alpha g_3 - 105 e_\alpha^2 g_2}{2^9.3.5^2.7} u^7
+ O(u^9).
\label{eqHalphenExpansions}
\end{eqnarray}

Full details on this symmetric notation of Halphen can be found in
Ref.~\cite{CCX}.

In the main text,
$(\alpha,\beta,\gamma)$ denotes any permutation of $(1,2,3)$.

\vfill \eject
\section*{Appendix B}
\label{sectionB}

The traveling waves of
(\ref{eqhNLS_continuous})
are defined by
the reduction
\begin{eqnarray}
& &
u=\sqrt{M(\xi)} e^{i (\varphi(\xi) - \omega t)},\ \xi=x-ct,
\label{eqhNLS_continuous_tw_reduction}
\end{eqnarray}
in which $M,\varphi$ are real functions
and $c,\omega$ real constants.

The two real functions $(M,\varphi)$ obey the sixth order
differential system
\begin{eqnarray}
& &
 c_2 \left\lbrack
-\frac{3 \varphi' M''}{2 M}
+\frac{3 \varphi' {M'}^2}{4 M^2}
-\frac{3 \varphi'' M'}{2 M}
-\varphi'''
+{\varphi'}^3
-\frac{3 q \varphi' M}{p}
    \right\rbrack
\nonumber \\ & &
+i c_2 \left\lbrack
\frac{M'''}{2 M}
-\frac{3 M' M''}{4 M^2}
+\frac{3 {M'}^3}{8 M^3}
-\frac{3 {\varphi'}^2 M'}{2 M}
+\frac{3 q M'}{2 p}
-3 \varphi' \varphi''
    \right\rbrack
\nonumber \\ & &
+c_1 \left\lbrack
 \frac{M''}{2 M} -\frac{{M'}^2}{4 M^2}- {\varphi'}^2 + \frac{q}{p} M
    \right\rbrack
+i c_1 \left\lbrack
  \varphi'' + \varphi' \frac{M'}{M}
    \right\rbrack
\nonumber \\ & &
+ \frac{c \varphi' + \omega}{p} -i \frac{c M'}{2 p M}
=0.
\label{eqhNLS_continuous_tw_reduction_ODE6complex}
\end{eqnarray}

All the first integrals of the system
(\ref{eqhNLS_continuous_tw_reduction_ODE6complex})
can be generated systematically
from the Lax pair \cite{AKNS} 
with the result,
\begin{eqnarray}
& &
K_1=c_2^2 \left(-\frac{M''}{3}
                + \frac{{M'}^2}{4 M}
                + {\varphi'}^2 M
                - \frac{q M^2}{2 p}\right)
-\frac{2}{3} c_1 c_2 \varphi' M
+ c_2 \frac{c M}{3 p},
\\ & &
K_2=c_2^2 \left(-\varphi' M''
                + \frac{\varphi' {M'}^2}{2 M}
                + \varphi'' M'
                + 2 {\varphi'}^3 M\right)
\nonumber \\ & & \phantom{xxxx}
+ c_1 c_2 \left(M''- \frac{{M'}^2}{M} + \frac{q M^2}{p} -4 {\varphi'}^2 M\right)
+ 2 c_1^2 \varphi' M
- (c_1 c + c_2 \omega) \frac{M}{p},
\\ & &
K_3=
c_2^2 \left(
\frac{{M''}^2}{4 M}
- \frac{{M'}^2 M''}{4 M^2}
- {\varphi'}^2 M''
+\frac{q}{p} M M''
+ \frac{{M'}^4}{16 M^3}
\right.
\nonumber \\ & & \phantom{xxxxxxx}
\left.
+ \frac{3 {\varphi'}^2 {M'}^2}{2 M}
- \frac{c {M'}^2}{2 p}
+ 2 \varphi' \varphi'' M'
+ {\varphi'}^4 M
+ \frac{q^2 M^3}{p^2}
\right)
\nonumber \\ & & \phantom{xxxx}
+ c_1 c_2 \left(
\varphi' M''
- \frac{3 \varphi' {M'}^2}{2 M}
- \varphi'' M'
-2 {\varphi'}^3 M
\right)
\nonumber \\ & & \phantom{xxxx}
-\frac{c_2 c}{p} M''
+\left(\frac{c_1^2}{4} + \frac{c_2 c}{2 p}\right)
 \left(\frac{{M'}^2}{M} + 4 {\varphi'}^2 M \right)
- 2 \frac{c_1 c + c_2 \omega}{p} \varphi' M
\nonumber \\ & & \phantom{xxxx}
+\frac{c_1^2 p - 4 c_2 c}{2 p^2} q M^2
+\frac{c_1 p \omega + c^2}{p^2} M.
\label{eqhNLS_continuous_tw_reduction_K1K2K3}
\end{eqnarray}

The associated spectral curve,
derived from the monodromy matrix $\mathcal{M}$,
\begin{eqnarray}
& &
\det(\mathcal{M}-\mu)=\mu^2-P(\lambda)=0,
\nonumber
\\
& &
P(\lambda)=
- \left(4 c_2 p \lambda^3 +2 c_1 p \lambda^2 + c \lambda - \frac{\omega}{2}\right)^2
- p q \left(6 K_1 \lambda^2 - K_2 \lambda + \frac{K_3}{2}\right),
\label{eqSpectralCurve}
\end{eqnarray}
is hyperelliptic and it has genus two.
Therefore the general solution is a singlevalued function of $\xi$
expressed with genus two hyperelliptic functions.
To our knowledge,
neither the separating variables nor the explicit expression
of this general solution are known.

When the polynomial $P(\lambda)$ has a double root,
the solution is elliptic 
and our goal here is to derive explicitly this elliptic
travelling wave solution.
It depends on the following arbitrary constants:
the origins $\xi_0,\varphi_0$ of $\xi$ and $\varphi$,
plus four constants out of $c,\omega,K_1,K_2,K_3$.
After this is achieved,
we want to
derive the corresponding solution of the discrete equation
(\ref{eqhNLS_discrete})
just by the natural discretization $x \to n h$ of this elliptic solution
of the continuum equation.


\vfill \eject

\end{document}